# Possible excitonic insulating phase in quantum confined Sb nanoflakes

Zhi Li[*,†,‡], Muhammad Nadeem[†, ‡], Zengji Yue[†, ‡], David Cortie[†, ‡], Michael Fuhrer[*,§,∥], Xiaolin Wang[*,†, ‡]

[†]ARC Centre of Excellence in Future Low-Energy Electronics Technologies (FLEET), University of Wollongong, Wollongong, NSW 2525, Australia

[‡]Institute for Superconducting and Electronic Materials (ISEM), Australian Institute for Innovative Materials (AIIM), University of Wollongong, Wollongong, NSW 2525, Australia

[§]School of Physics and Astronomy, Monash University, Victoria 3800, Australia

[∥]ARC Centre of Excellence in Future Low-Energy Electronics Technologies (FLEET), Monash University, Victoria 3800, Australia

## ABSTRACT

In the 1960s, it was proposed that, in small indirect band gap materials, excitons can spontaneously form because the density of carriers is too low to screen the attractive Coulomb interaction between electrons and holes. The result is a novel strongly interacting insulating phase known as an excitonic insulator. Here, we employ scanning tunnelling microscopy (STM) and spectroscopy (STS) to show that the enhanced Coulomb interaction in quantum confined elemental Sb nanoflakes drives the system to the excitonic insulator state. The unique feature of the excitonic insulator, a charge density wave (CDW) without periodic lattice distortion (PLD), is directly observed. Furthermore, STS shows a gap induced by the CDW near the Fermi surface. Our observations suggest that the Sb(110) nanoflake is an excitonic insulator.

The discovery of new phases of matter is one of the major goals of condensed matter physics and is important for developing new technologies. In the insulator family, the first member is the band insulator (Figure 1a), which is well explained by the band theory of non-interacting electrons. In a band insulator, the insulating state is induced by fully occupied orbitals, and is characterized by a finite band gap ($E_g$) between valence and conduction bands. Besides band insulators, other insulating states may arise through the effects of electron-electron interactions or disorder coupled with quantum interference. For example, in Mott insulators, electrons are localized by the strong repulsive Coulomb interaction, which prevents electron hopping. In Anderson insulators, electrons are localized by quantum interference. The recent understanding of topological phases of electronic matter has added topological insulators[1,2], which have a gap in the bulk but gapless conducting states[3] at the surface/edge (Figure 1b) due to band inversion as a result of spin-orbital coupling.

An intriguing insulating state, named the excitonic insulator, was predicted decades before[4-6], where bosonic particles rather than electrons determine the physical properties (Figure 1c). Excitons, bosonic strongly bound pairs of electrons and holes, are formed through the attractive electron-hole Coulomb interaction, lowering the system energy by the value of the binding energy ($E_b$). If such excitons could form spontaneously, the result would be an excitonic insulator phase, first predicted to be a superfluid phase, because of its bosonic ground state[7,8]. Excitons can also form a solid-like structure, called crystallized excitonium[9].

In semiconductors or insulators, the formation of an exciton requires overcoming the band-gap energy $E_g$ needed to create an electron-hole pair. The spontaneous formation of excitons demands ($E_b > E_g$). However, $E_g$ is usually much larger than $E_b$ in semiconductors and insulators preventing spontaneous exciton formation. This presents two potential strategies in searching for potential excitonic insulators. The first strategy is to focus on materials with nearly zero band overlap ($E_g \approx 0$) or zero gap materials[3], in which even a small $E_b$ can bind electrons and holes, forming excitons spontaneously (Figure 1c). This strategy has produced a number of interesting candidate materials, such as 1T-TiSe$_2$[8] and InAs/GaSb quantum wells[10]. Here, we adopt a second strategy, exploiting the large Coulomb interaction by quantum confinement to enhance $E_b$ and thus overcome $E_g$. The nature of the Coulomb screening effect is the movement of free electrons at the Fermi surface compensated the Coulomb potential, which is described in the Thomas–Fermi screening theory. Therefore, when a material length scale is comparable to the wavelength of free electrons at the Fermi surface (Fermi wavelength) the strength of Coulomb screening will be severely reduced. This technique has been widely applied in photonic semiconductor quantum wells devices[11]. The width of quantum wells is made comparable to the Fermi wavelength of the two-dimensional electron gas, which is around 40 nm in GaAs/Al$_x$Ga$_{1-x}$As[12]. Therefore, the binding energy of excitons is enhanced greatly. In noble metals, the Fermi wavelength is small, which is usually less than 1 nm[12] and the Coulomb interaction is very weak. In semimetals, the small Fermi surface leads to a large Fermi wavelength and stronger Coulomb interaction. For example, Sb has a Fermi wavelength as large as 5.5 nm[12]. Therefore, a significant enhancement of Coulomb interaction is expected in Sb nanoflakes with a thickness of few nanometers. One direct consequence of exciton formation in a semimetal is the gap opening at the Fermi surface and the modulation of charge density in real space, which are also characteristic features of the Peierls instability-induced CDW. Recent reports have identified several materials, including TmSe$_{1-x}$Te$_x$, Ta$_2$NiSe$_5$, and TiSe$_2$, as candidate excitonic insulators by studying their CDW[8,10,13-15]. Among them, TiSe$_2$ is believed to be the most promising candidate excitonic insulator, because its small indirect band gap matches well with the criterion required for excitonic insulators[8,15-17]. Nevertheless, the significant PLD in TiSe$_2$ makes it hard to distinguish it from Peierls phase, leading to a debate on the nature of the insulating phase[18-20]. In both Peierls and excitonic insulator phase, the CDW phase arises from the nesting of the Fermi surfaces of electron and hole at a different region of the Brillouin zones. In Peierls phases, PLD forms to decrease the system energy and then drives a periodical charge distribution. Therefore, PLD and CDW are critical identities for Peierls phase. In contrast, in the excitonic insulator phase, the system energy is lowered by the formation of excitons. The lattice is able to find a lower energy state by responding to the electronic charge redistribution through the electron-phonon coupling. Nevertheless, due to the small electron-phonon coupling constant of Sb[21], the PLD will be very small. Thus, an observation of a CDW phase without PLD will unambiguously identify the excitonic insulator phase.

Sb has long been considered a candidate excitonic insulator because of its semimetal nature with very small band overlap[6-8]. However, the exciton binding energy in *bulk* Sb was believed to be very small[6], making it almost impossible to observe the excitonic insulator phase. Up to now, no CDW or other signature of an excitonic insulator has been reported in Sb. The advent of the quantum confinement materials, in which strong Coulomb interactions can lead to enormous exciton binding energies, provides us with a new way to realize an excitonic insulator. In quantum-well structures, the exciton binding energy is enhanced significantly due to weak Coulomb screening[11]. In this work, we demonstrate that Sb(110) nanoflake with a thickness comparable to the Fermi wavelength is an excitonic insulator by direct observation of a CDW without PLD and gap opening at the Fermi level.

Molecular beam epitaxy (MBE) was used to prepare polycrystalline Sb nanoparticles with sizes of few tens of nanometers on graphene/SiC(0001) substrates (Figure S1 in the Supporting Information). Before depositing Sb, SiC(0001) substrates were annealed at 1350°C for 10 min under ultra-high vacuum (UHV, $< 1 \times 10^{-10}$ Torr) to obtain double-layer graphene. The deposition flux of Sb was 1 monolayer per minute (ML/min). The temperature of the Graphene/SiC substrate was kept at room temperature during deposition. The STM and STS measurements were carried out using a low-temperature UHV scanning tunnelling microscopy system (LT-UHV-STM) at 77 K. The *dI/dV* spectroscopy was acquired by switching off the feedback loop and keeping a constant tip-sample distance. The modulation amplitude of 10 mV at 673 Hz was applied when conducting STS measurements. To investigate the Coulomb confinement effect on Sb, we prepared nanoflakes by using STM tip to "cleave" nanoparticles in-situ. As shown in Figure S1 in the Supporting Information, by applying controllable STM tip bias pulses, many Sb nanoflakes are separated from the polycrystalline Sb nanoparticles. The separated nanoflakes include Sb(110), Sb(111) nanoflakes (in rhombohedral index notation)[22], and recently discovered α-animonene[23]. Figure 2a shows a three-dimensional STM image of a Sb(110) nanoflake with a thickness of 5 nm on a graphene substrate. The Sb nanoflake shows a pseudo-square surface lattice (Figure 2b and structure model in Figure S2), where $a_0$ = 0.45 nm and $b_0$ = 0.41 nm are the lattice constants for this pseudo-square lattice and the step height is 0.31 nm. The pseudo-square surface lattice, step height, and lattice constants are all in good agreement with the previously reported bulk truncated Sb(110) surface[22]. Here, second layer Sb atoms, which are slightly below the first layer atoms, are also observed in STM images. The typical large-scale *dI/dV* spectrum is V-shaped with a very low density of states (DOS) at the Fermi surface (Figure 2c), consistent with band structure calculations on few-layer Sb thin films[24], where the DOS minimum near the Fermi surface is the consequence of a small band overlap between the conduction band and the valence band.

Figures 2d and 2e show an atomic resolution STM image and the simultaneously obtained DOS mapping within the gap energy (20 mV) on this nanoflake. The STM image in Figure 2d shows a well-defined pseudo-square lattice without any signature of a PLD. DOS mapping at 20 meV (Figure 2e) shows stripe-like ordering, where bright spots denote a higher DOS. The height profile along the black dashed line in Figure 2d shows corrugations with a periodicity of one unit cell (upper part of Figure 2f). In contrast, the DOS profile along the red dashed line in Figure 2e shows corrugations with a periodicity of two unit cells (lower part of Figure 2f), suggesting a 1 × 2 striped DOS ordering without lattice distortion. In materials without strong electron correlation, stripe ordering is usually observed in moiré patterns, quasi-particle interference (QPI) from step edges, and CDWs. The striped DOS ordering observed here is not a moiré pattern induced by lattice mismatch between Sb nanoflakes and the graphene substrate. A moiré pattern should show significant lattice distortion, which is not observed here. The striped DOS ordering is also not QPI induced by scattering from step edges. The step edge induced QPI should be stronger at the step edge. In contrast, the DOS order observed here is stronger away from the edge (Figure 2d and 2e). The periodic DOS modulation without lattice distortion (electron-driven CDW) observed here is the unique signature of an excitonic insulator.

Because lattice distortion is a common origin for a CDW, to distinguish an excitonic insulator from Peierls insulators, any possible periodical lattice distortion must be investigated carefully. In Figure 3, we carried out a systematic fast Fourier transform (FFT) investigation on spontaneously obtained atomic resolution STM image (Figure 3a) and DOS mapping (Figure 3f). The image shows twin domain boundaries, which is a natural consequence of the pseudo-square lattice (Figure S3). The 1 × 2 striped DOS ordering is perpendicular to each other in adjunct domains (highlighted by black and red ellipses in Figure 3f). To highlight the 1 × 2 striped ordering, we filtered signals with a wavelength larger than the 1 × 2 striped ordering from Figure 3a and Figure 3f. After this filter was applied, the DOS image (Figure 3h) shows more clear evidence of 1 × 2 striped ordering. Instead of features corresponding to the 1 × 2 striped ordering, the STM image (Figure 3c) shows only a faint feature of 1 × 1 lattice and terraces. We also did reverse FFT images by only keep information around 1 × 2 and 2 × 1 spot. The filtered DOS mappings show that vertical stripes mainly locate at the right lower part of the DOS mapping (Figure 3i) and the lateral stripes mainly locate at the upper left part (Figure 3j), which are consistent with the original DOS mapping (Figure 3f). In contrast, the lateral and vertical stripes distributed almost evenly in the filtered STM images (Figure 3d and 3e). Therefore, our results confirm that the CDW observed here is not accompanied by any periodic lattice distortion, strongly indicating the characteristic feature of an excitonic insulator.

Another unique signature of excitonic insulators is the opening of a gap in the DOS spectrum[6,17]. The typical small-scale *dI/dV* spectrum of Sb nanoflakes (Figure 4a) shows a 70 meV gap near the Fermi level. The non-zero DOS inside the gap may due to the partly gap opening at the Fermi surface. DOS mappings indicate that the opening of this gap is driven by the excitonic insulating state. Figure 4c-4f shows DOS mappings of the region shown in Figure 4b at various energies (50 mV, 20 mV, -20 mV, and -50 mV). The stripe charge order is very strong in mappings at the center of the gap, with strong contrast between high DOS and low DOS regions (Figure 4e). Away from the gap center, the DOS mapping shows weaker stripe charge ordering (Figure 4e), which is evidenced by a decreased contrast between high and low DOS regions. DOS mapping at the very edge of the gap (Figure 4f) shows only faint DOS modulation. However, the periodicity of the CDW did not change on varying the bias voltage. From the theory of excitonic insulator, the periodicity of the CDW is determined by the $q$-vector connecting electron and hole pockets and will not change with bias voltage. This strongly suggests that the gap and CDW have the same origin and further exclude the likelihood of a QPI effect. Furthermore, high DOS regions (bright regions) are not exactly located on the top of the topmost Sb atoms (Figure S4). The discrepancy between spatial variations in DOS and atom position is in strongly contrast with the Peierls transition induced CDW, where charge modulation appears on the topmost atoms due to lattice distortion. Since the CDW induced by the excitonic insulating state stems from instability of the band structure rather than instability in the real-space ionic lattice, DOS modulation will follow orbitals that contribute to the formation of excitons rather than the positions of the nuclei. The observation of a CDW without PLD and gap opening at the Fermi surface clearly suggests that the Sb(110) nanoflake is an excitonic insulator.

In summary, a CDW without lattice distortion and band gap opening at the Fermi surface, the two unique signatures of an excitonic insulating phase, are observed in Sb(110) nanoflakes. Our findings provide a novel strategy to search for more excitonic insulators. Further studies, e.g. temperature and pressure dependence experiments, are required to fully understand the rich physics of the excitonic insulator.

## SUPPORTING INFORMATION

**STM tip in-situ cleaving technique and varies Sb phases, additional structure information, and detailed analysis of the spatial variations in DOS and atom position.**

## AUTHOR INFORMATION

**Corresponding Author**
*xiaolin@uow.edu.au; *michael.fuhrer@monash.edu; *zhili@uow.edu.au

## ACKNOWLEDGMENTS

The authors acknowledge support from the Australian Research Council (ARC) through the ARC Centre of Excellence in Future Low-Energy Electronics Technologies (FLEET, CE170100039). XLW acknowledges support from an ARC Professorial Future Fellowship project (FT130100778). Z. Li acknowledges support by ARC Discovery Projects (DE190100219, DP160101474, DP170104116) and the University of Wollongong through the Vice Chancellor's Postdoctoral Research Fellowship Scheme. MSF acknowledges the support of an ARC Laureate Fellowship FL120100038. The STM instruments and consumables are partly supported by DP170101467, LE100100081 and LE110100099. We thank Professor Chao Zhang for helpful discussion.

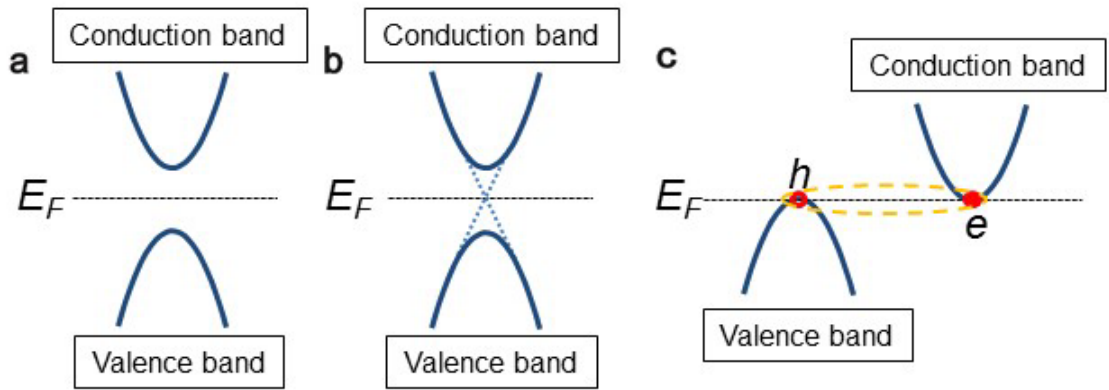

**Figure 1.** Three types of insulators. (a) Insulators with a band gap at the Fermi surface, including band insulators, Mott insulators and Anderson insulators. (b) Topological insulators, where there are topologically protected surface states inside the band gap. (c) Excitonic insulators, where electrons and holes are bound together and form excitons at the Fermi surface. The ideal case occurs where the conduction band and the valence band have a very small overlap at the Fermi level.

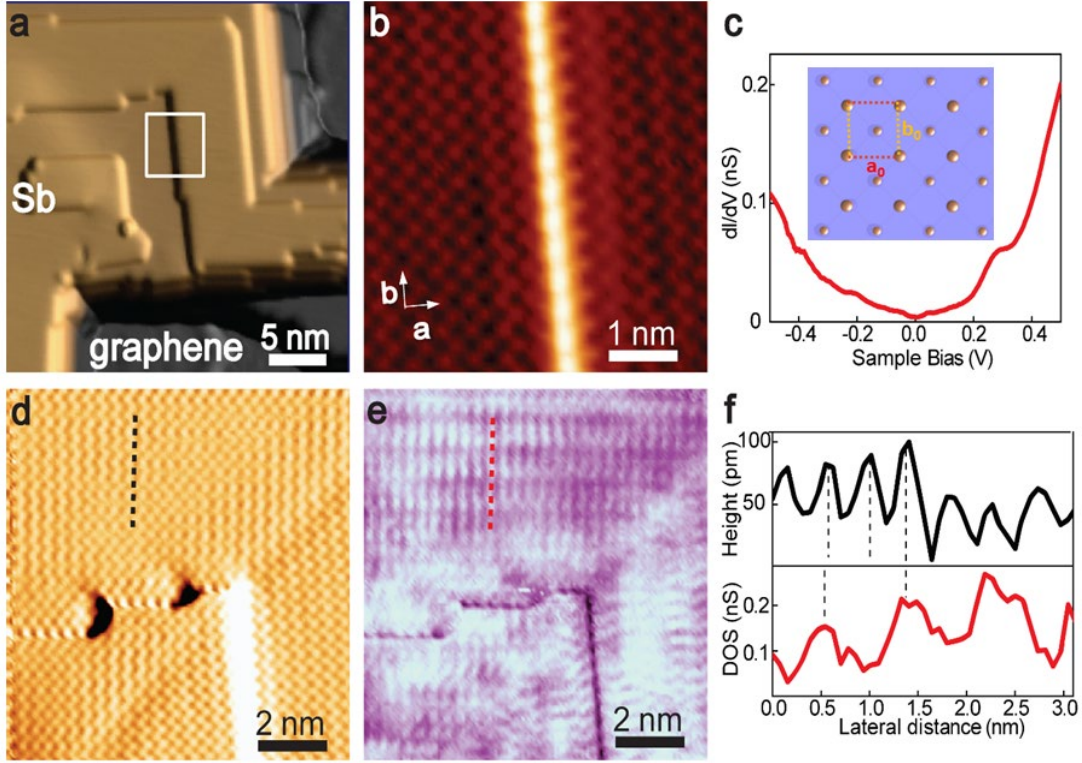

**Figure 2.** CDW without PLD on Sb nanoflakes. (a) STM image of an Sb nanoflake (30 nm × 30 nm, $V_s$ = 3 V, $I_t$ = 50 pA) with thickness of around 5 nm. (b) STM image of region enclosed by the white square in (a) shows a pseudo-square lattice with $a_0$ = 0.45 nm and $b_0$ = 0.41 nm, which is a perfect Sb surface without reconstruction (5 nm × 5 nm, $V_s$ = 20 mV, $I_t$ = 100 pA). (c) Typical STS curve on Sb nanoflakes shows a V-shaped feature near the Fermi surface, which is a typical feature of a semimetal with a very small indirect negative band gap ($V_s$ = 500 mV, $I_t$ = 100 pA). Inset shows the lattice structure. (d) STM image with atomic resolution of an Sb(110) nanoflake (10 nm × 10 nm, $V_s$ = 20 mV, $I_t$ = 100 pA). (e) Simultaneously obtained *dI/dV* mapping shows DOS modulation along the *a* direction. Brighter color denotes a higher DOS. (f) The top part shows a height profile along the black dashed line in (d). The lower part shows a line profile of the DOS (red dashed line in (e)) with a $2a_0$ periodic modulation.

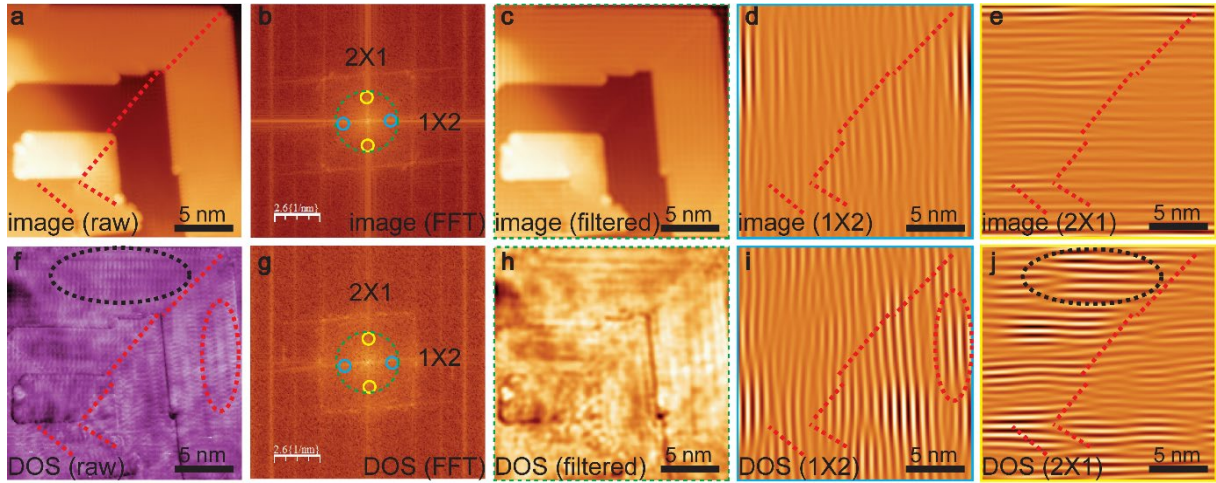

**Figure 3.** FFT analysis showing the absence of PLD. (a) and (f) are topographic image and simultaneously obtained DOS mapping of a Sb nanoflake, respectively (20 nm × 20 nm, $V_s$ = 20 mV, $I_t$ = 50 pA). The domain boundaries are marked by red dashed lines. The FFT image of the (a) and (d) are shown in (b) and (g), respectively. The 1 × 2 and 2 × 1 spots are denoted by blue and yellow circles, respectively. (c) and (h) show filtered images of (a) and (f), respectively, by excluding the long wavelength components, the region outside the green circles in (b) and (g). (d) and (e) are filtered image of (a) showing the spatial distribution of region around the 1 × 2 and 2 × 1 spot. (i) and (j) are filtered DOS mapping of (f) showing the spatial distribution of region around the 1 × 2 and 2 × 1 spots. The domain boundaries in (a), (d), (e), (f), (i) and (j) are marked by red dashed lines. Higher height or DOS is denoted by brighter colour in each image.

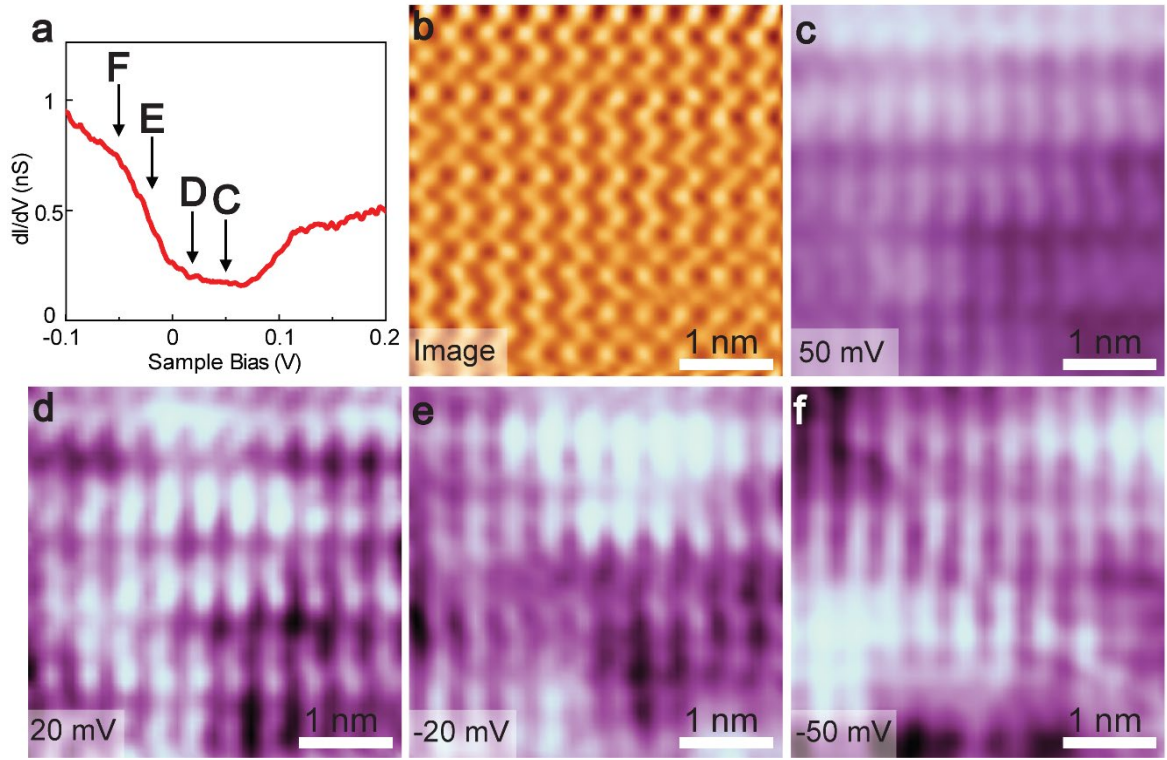

**Figure 4.** Bias dependence of CDW. (a) Typical small-scale $dI/dV$ vs. sample bias curve shows a gap opening near the Fermi surface ($V_s$ = 0.2 V, $I_t$ = 100 pA). (b) Atomic resolution image of a Sb nanoflake with a thickness of 5 nm (4 nm × 4 nm, $V_s$ = 20 mV, $I_t$ = 100 pA). Images (c), (d), (e), and (f) show $dI/dV$ mappings of the region shown in (b) at 50 mV, 20 mV, -20 mV, and -50 mV, respectively. Biases are marked at the lower left corner of each mapping image and on (a). Imaging size of panels (b-f) is 4 nm.

TOC Graphic:

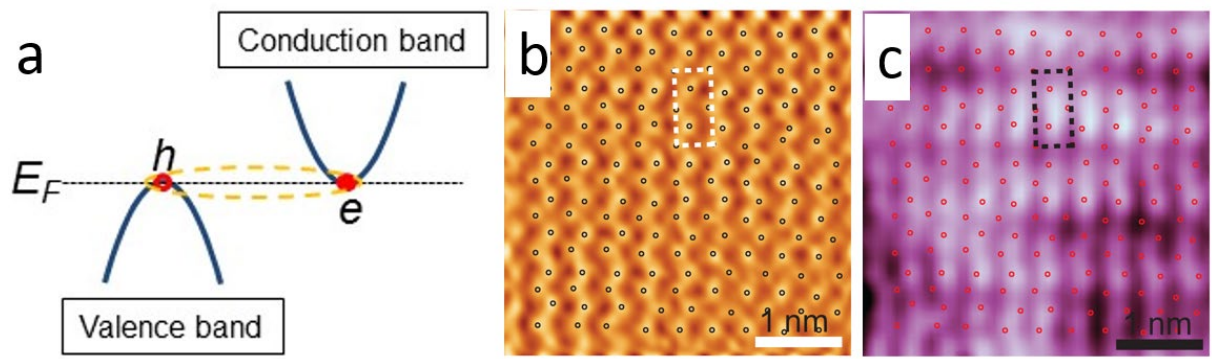

We employ scanning tunnelling microscopy (STM) and spectroscopy (STS) to study the electronic states of elemental Sb(110) nanoflakes. The unique feature of the excitonic insulator, a charge density wave (CDW) without periodic lattice distortion (PLD), is directly observed.